\documentclass[twocolumn,showpacs,superscriptaddress,prb,aps,floatfix]{revtex4}

\usepackage{graphicx}
\usepackage{rotating}
\usepackage{amsmath}

\newcommand{\bab}{$\rm BaBiO_3$}
\newcommand{\bak}{$\rm Ba_{1-x}K_xBiO_3$}

\begin{document}

\title{Structural, vibrational and quasiparticle properties of the Peierls semiconductor $\rm BaBiO_3$: 
a  hybrid functional and self-consistent GW+vertex-corrections study}

\author{C.~Franchini}
\affiliation{
Faculty of Physics, Universit\"at Wien and Center for Computational Materials Science,  Sensengasse 8, A-1090 Wien, Austria}

\author{A. Sanna}
\affiliation{
Department of Physics, University of Cagliari and Sardinian Laboratory for Computational Materials 
Science SLACS (INFM-CNR), Cittadella Universitaria, I-09024 Monserrato (Ca), Italy}

\author{M.~Marsman}
\affiliation{
Faculty of Physics, Universit\"at Wien and Center for Computational Materials Science,  Sensengasse 8, A-1090 Wien, Austria}

\author{G.~Kresse}
\affiliation{
Faculty of Physics, Universit\"at Wien and Center for Computational Materials Science,  Sensengasse 8, A-1090 Wien, Austria}

\date{\today}

\begin{abstract}
{
$\rm BaBiO_3$ is characterized by a charge disproportionation with half of the
Bi atoms possessing a valence 3+ and half a valence 5+.  
Because of selfinteraction errors, local and semi-local density functionals fail to describe the 
charge disproportionation quantitatively, yielding a too small structural
distortion and no band gap. Using hybrid functionals 
we obtain a satisfactory description of the structural, electronic, optical, and 
vibrational  properties of $\rm BaBiO_3$. The results obtained       
using GW (Green's function G and screened Coulomb potential W) based schemes 
on top of hybrid functionals, including fully selfconsistent GW calculations
with vertex corrections in the dielectric screening, qualitatively confirm the HSE picture
but a systematic overestimation of the bandgap by about 0.4 eV is observed.   
} 
\end{abstract}

\maketitle

%================================
\section{Introduction}
\label{sec1}
%===============================

\begin{figure}
\includegraphics[clip,width=0.45\textwidth]{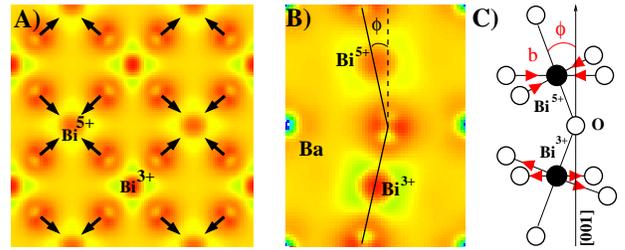}
\caption{(Color online) Self-consistent-GW charge density plots showing
(a) the breathing distortion in the (100) plane and (b) the tilting  instability
projected in the (001) plane. $\rm Bi^{5+}$ (Bi1) and $\rm Bi^{3+}$ (Bi2) indicate
the two inequivalent bismuth ions, whereas the arrows indicate the breathing displacement
of the oxygen atoms. (c) Schematic view of the \bab\ unit cell showing the $\rm BiO_6$ tilted
octahedra. Color coding: darker (red) areas indicates a high value of the charge density.}
\label{bt}
\end{figure}

The Peierls semiconductor \bab, parent compound of the high-$\rm T_c$ superconductor \bak,
has long been of theoretical and experimental interest due to its distinct negative-U nature.
This behavior can be understood through the concept of "forbidden valence" suggesting that in \bab\ 
bismuth atoms appear in two different valences (${5+}$ and ${3+}$), but {\em skip} the forbidden formal
valence in-between (${4+}$). 
The ``valence-skipper'' bismuth atoms hence disproportionate, i.e. every second Bi atom
donates all 5 valence electrons to oxygen, whereas the other Bi atom retains two
6s electrons and only donates the three 6p electrons to the oxygen atoms. 
Such a behavior is usually explained by the concept of a negative U which
invalidates the usual positive Coulomb repulsion between two electrons\cite{harrison}.
The charge disproportionation goes in hand with significant structural changes from the 
ideal cubic perovskite crystal towards a distorted monoclinic structure
characterized by a short-ranged charge density wave (CDW) state formed by alternating breathing-in
and breathing-out distortions of oxygen octahedra around inequivalent
$\rm Bi^{5+}$ (Bi1) and $\rm Bi^{3+}$ (Bi2) ions (Fig. \ref{bt}).
As a consequence of this charge ordering, the formally expected metallic state for the cubic 
perovskite $\rm BaBi^{4+}O_3$ is replaced by an insulating regime characterized
by a large direct CDW optical response $\rm E_d\approx$ 2.0 eV\cite{direct}
and an indirect optical transition $\rm E_i$.
Although \bab\ has been the subject of numerous optical investigations, dissenting opinions
were reported on the value of the indirect gap $\rm E_i$, for which conflicting measurements
are reported in literature, ranging from 0.2 eV\cite{expgap1} and 1.1 eV\cite{kunc}. 

Despite the apparent absence of strong electron localization usually
associated with $d$ and $f$ electrons, standard semilocal density
functionals exhibit severe problems in describing the basic properties of this material, 
yielding much too small or even negative band gaps\cite{kunc,rabe}, thus preventing a detailed description
of the electronic properties. In a recent paper, we have explained the polaron-mediated insulator-to-metal
transition experimentally observed in K-doped $\rm BaBiO_3$ by means of hybrid density functional theory. 
We have shown that semilocal functionals are incapable (i) to disclose the underlying physics of this type of
electronic transition
and, more generally, (ii) to account for "long range" Peierls-like distortions 
in doped \bab\cite{cfgk}. 
The objective of the present study is to apply the screened exchange hybrid density 
functional theory (DFT) Heyd-Scuseria-Ernzerhof (HSE)
functional\cite{hse} to the ground state structural, electronic and vibrational properties of \bab,
and to investigate to which extent many-body effects may alter the ground state and quasi-particle (QP)
properties of \bab. We address the latter issue by adopting the GW approach. These calculations
essentially confirm the picture emerging from the hybrid functional.
Details of our computational approach are described in
the methodological section, whereas the results are presented and discussed in Sec.\ref{sec:3}.

\begin{table*}[ht]
\caption{Collection of calculated quantities along with the available experimental data.
The abbreviations $V$ (volume), $b$ (breathing distortion), $\phi$ (tilting distortion), $\beta$ (monoclinic angle),
$E_d$ (direct gap), $E_i$ (indirect gap) and${\epsilon_\infty}$ (macroscopic dielectric constant) are described in 
the text and, partially,  in Fig. \ref{bt} (c).
$^{a}$ Ref.\onlinecite{expgeo},
$^{b}$ Ref.\onlinecite{direct}
$^{c}$ Ref.\onlinecite{expgap1}
$^{d}$ Ref.\onlinecite{expgap2}
$^{e}$ Ref.\onlinecite{expgap3}
$^{f}$ Ref.\onlinecite{kunc},
$^{g}$ Ref.\onlinecite{expeps1},
$^{h}$ Ref.\onlinecite{expeps2},
}
\vspace{0.3cm}
\begin{ruledtabular}
\begin{tabular}{cccccccccc}
             & PBE& PBEsol  & HSE & $\rm G_0W_0$ & $\rm GW_0^{TCTC}$& $\rm scQPGW^{TCTC}$ &LMTO-LDA & PAW-LDA &  Expt.     \\
&\multicolumn{6}{c}{(This work)}&(Ref.\onlinecite{kunc})&(Ref.\onlinecite{rabe})&   \\
                     &      &&     &      &      &    &       &      &               \\
$V$ (\AA$^{\rm 3}$)  & 85.76&85.94&82.10&   -  &   -  & -  & 80.41 & 81.04&  82.21 $^{a}$ \\
$b$ (\AA)            & 0.07 &0.075&0.09 &   -  &   -  & -  &0.11   & 0.04 &  0.085 $^{a}$ \\
$\phi$ ($^{\circ}$)  & 12.1 &12.2&11.9 &   -  &   -  & -  &9.6    &  -   &  11.2 $^{a}$  \\
$\beta$ ($^{\circ}$) & 90.16&90.14&90.24&   -  &   -  & -  &  -    &  -   &  90.17 $^{a}$ \\
                     &      &&     &      &      &    &       &      &               \\
$E_d$ (eV)           & 1.22 &1.32&2.07 & 2.62 & 2.39 &2.45&1.1    &  -   &  2.0 $^{b}$   \\
$E_i$ (eV)           & 0.0  &0.0&0.84 & 1.32 & 1.15 &1.28&0.1    & 0.0  &  0.2 $^{c}$, 0.5 $^{d}$, 0.8 $^{e}$, 1.1 $^{f}$
                                  % 1.4 G5W0
\\
                     &      &&     &      &      &    &       &      &  \\
${\epsilon_\infty}$  & 12.2 &&6.0  &      & 7.0  & 7.2& -     &   -  &  5 $^{g}$, 7 $^{h}$     \\
\end{tabular}
\end{ruledtabular}
\label{data}
\end{table*}

%===============================
\section{Methodology}
\label{sec:2}
%===============================

We have used the projector augmented wave method\cite{paw1,paw2} based Vienna {\em ab initio}
simulation package (VASP)\cite{vasp} employing the generalized gradient approximation scheme of
Perdew {\em et al.}\cite{gga} (PBE), the HSE hybrid functional,\cite{hsevasp} and the GW
formalism\cite{gwvasp06,gwvasp07}.
The HSE functional is constructed by mixing 25\% of the exact
Hartee-Fock (HF) exchange with 75\% of the Perdew-Burke-Ernzerhof (PBE) exchange functional\cite{hse}.
The resulting expression for the exchange (x) and correlation (c) energy is
\begin{equation}
E_{xc}^{\rm HSE03} = E_{x}^{\rm PBE} -  \frac{1}{4}E_{x}^{\rm PBE,sr,\mu} + \frac{1}{4}E_{x}^{\rm HF,sr,\mu} + E_{c}^{\rm PBE},
\end{equation}
in which the parameter $\mu$ controls the decomposition of the Coulomb kernel
into short-range (sr) and long-range  contributions to the exchange.
$E_{x}^{\rm HF,sr,\mu}$ is the exact exchange screened at very long range, and
$E_{x}^{\rm PBE,sr,\mu}$ is the corresponding short range exchange functional constructed
using the PBE exchange hole. 
According to the HSE06 recipe, we set $\mu=0.2$ $\rm \AA^{-1}$.\cite{hse6}
This implies that non-local exchange is rather long ranged extending over
several nearest neighbor shells.
The HSE approach has been used to calculate the starting orbitals for different GW-based schemes
of increasing complexity and precision in the evaluation of the self-energy $\Sigma = iGW$.
The screened Coulomb kernel $W = \epsilon^{-1}v$ which enters in the previous expression requires 
the computation of the frequency dependent dielectric matrix $\epsilon^{-1} = 1+v\chi$, where 
the polarizability $\chi$ is given by the Dyson equation
\begin{equation}
\chi = [1-\chi_0(v+f_{xc})]^{-1}\chi_0
\label{dyson}
\end{equation}
Here, $\chi_0$ indicates the independent particle polarizability, whereas $f_{xc}$ represents
the effective non-local exchange correlation kernel which describes the many-body interaction
between electrons and holes (vertex corrections). We have adopted three different GW strategies:

(i) The widely used single shot (i.e. non-self-consistent) $\rm G_0W_0$ approximation, where 
$\rm G_0$ and $\rm W_0$ are calculated using HSE eigenvalues and eigenfunctions within
the so called random phase approximation (RPA) (i.e. excluding any vertex corrections, $f_{xc}$=0).\cite{Louie} 
The VASP implementation is detailed in Refs. \onlinecite{gwvasp06} and \onlinecite{gwvasp07}. Using
HSE orbitals, this approach usually slightly overestimates
the band gap, but is very efficient.\cite{fuchs07}

(ii) A partially self-consistent procedure consisting of an update of the eigenvalues in the Green's 
function G combined with HSE screening properties W$_0$ ($\rm GW_0$). Within this scheme
we have included electrostatic electron-hole interactions
in the computation of W using time-dependent HSE (TD-HSE) through an effective
non-local frequency-independent kernel $f_{xc}$.
This specific GW scheme, usually denoted as $\rm GW_0^{TCTC}$ (test-charge/test-charge $\rm GW_0$), 
is believed to remedy most of the problems of single shot $\rm G_0W_0$ and is considered to be a convenient 
(less time-consuming) and trustable (experimental precision) alternative to fully self-consistent GW. 
The approach was found to yield excellent QP band gaps in 
simple $sp$ bonded semiconductors, if and only if the electrostatic interactions between
electrons and holes were taken into account.\cite{Paier08}
The electrostatic electron-hole interaction is included using an effective many-electron
exchange correlation kernel constructed using the HSE functional (Nano-quanta kernel 
determined to mimic TD-HSE, for details we refer to Refs. \onlinecite{vc,sottile,adragna03,bruneval05}).

(iii) Finally, we employ the self-consistent quasiparticle GW ($\rm scQPGW^{TCTC}$), 
following the original ideas
of Faleev, Schilgaarde, and Kotani\cite{scqpgw,schigw}, with the inclusion of vertex corrections
adopting the formalism of Reining {\em et al.}\cite{vc,sottile,adragna03,bruneval05}
[Nano-quanta kernel constructed to mimic the Bethe-Salpeter equation (BSE)]. 
This approach
is rather time-consuming but entirely free of any dependence on the starting orbitals.
For $sp$ bonded materials, it yields band gaps within a few percent of experiment,
if the electrostatic electron-hole interaction is included using
a many-electron exchange correlation kernel $f_{xc}$.\cite{scgwvasp}

The simulated unit cell, containing 10 atoms, has been modeled using a 
4$\times$4$\times$4 {\em k}-point grid and an energy cutoff of 400 eV,
reduced to 300 eV within $\rm GW$ calculations. To calculate the frequency-dependent dielectric matrix 
the inclusion of 400 empty bands was sufficient to obtain well converged screening properties and QP eigenvalues.

%===============================
\section{Results and Discussion}
\label{sec:3}
%===============================

\subsection{Structure}

The low temperature phase of \bab\ is monoclinic and derives from the primitive 
cubic perovskite high temperature cell by simultaneous breathing ($b$) and tilting ($\phi$) distortions of the
octahedra\cite{expgeo} (Fig. \ref{bt}). 
Unlike previous studies, we have performed a full structural relaxation
by allowing for both $b$ and $\phi$ distortions as well as for volume (V) and shape ($\beta$, the monoclinic angle) optimization.
The resulting minimized geometries listed in Tab. \ref{data} reveal that the structural
properties are sensitive to the applied functional and are generally better reproduced within
HSE. 
In particular, the PBE functional seriously overestimates the volume ($\approx$ 4 \%), which is in fact 
well described using HSE. Also, 
the breathing displacement, which is responsible for the opening of the
CDW gap\cite{kunc}, is better described using the hybrid functional. The minimum HSE Bi-O
bond length splitting $b$ = 0.09 \AA\ matches exactly the experimental value. 
The volume and, to a lesser extent, the breathing displacement does not improve by using a revised version
of the PBE functional designed to yield improved equilibrium volumes, the so called
PBEsol functional\cite{pbesol} (see Tab. \ref{data}).

\subsection{ Electronic properties}

The approximations made in the exchange-correlation term,
have an even more drastic impact on the predicted  electronic and dielectric properties.
In line with past findings, PBE predicts a zero-gap system ($E_i^{PBE}$ = 0.0 eV) with a much to small
direct band gap $E_d$ ($E_d^{PBE}$ = 1.22 eV).
Although the origin of this failure has often been related to
the deficient description of the structural instabilities at the PBE level, we found that
the values $E_i^{PBE}$ and $E_d^{PBE}$ computed adopting the HSE and experimental geometries 
are only 0.05 meV larger than those calculated at the PBE structural minimum, 
and therefore still in conflict with the experimental measurements.

\begin{figure*}
\includegraphics[clip,width=0.7\textwidth]{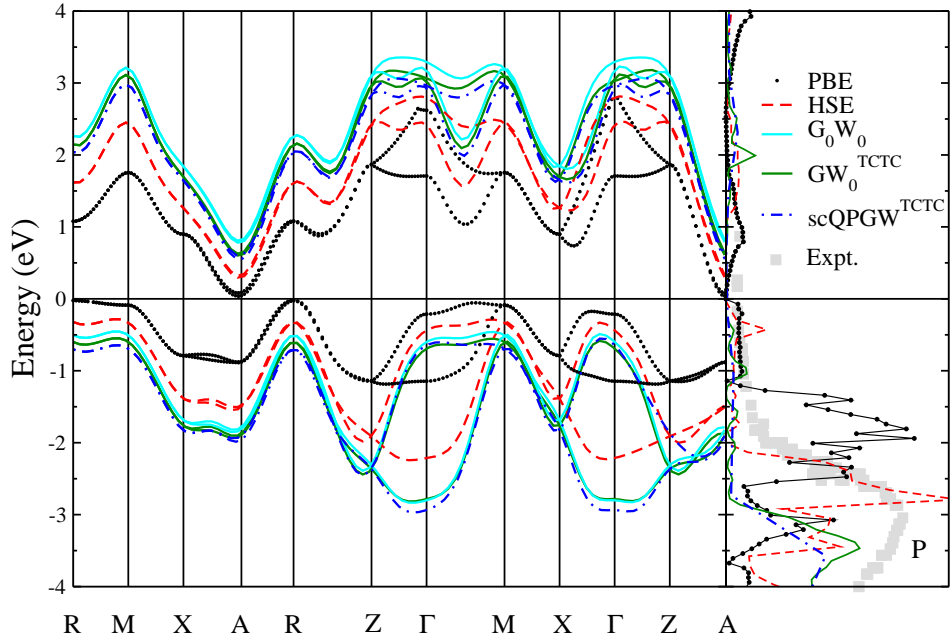}
\caption{(Color online) Calculated (left) band structure of monoclinic \bab\ within PBE, HSE $\rm G_0W_0$, $\rm GW_0^{TCTC}$ and $\rm scQPGW^{TCTC}$
and (right) corresponding total density of states.
$E_i$ is opened between the A and R points of our modeled supercell, which correspond to the W and L 
points of the simple monoclinic cell. P refer to the measured main photoemission peak of the valence band\cite{peak}.
}
\label{band}
\end{figure*}

In order to capture the insulating character of \bab\, we have adopted the HSE functional and the GW
method at the three levels described in the methodological section.
The computed band structure is shown in Fig.\ref{band}, where we plot the most relevant bands 
near the Fermi level ($E_F$=0) for PBE, HSE, $\rm G_0W_0$, $\rm GW_0^{TCTC}$ and $\rm scQPGW^{TCTC}$.
We observe that the pragmatic inclusion of 25\% non-local Hartree-Fock exchange
within the HSE approach shifts apart the bands near $E_F$ and increases both
$E_i$ (0.84 eV) and $E_d$ (2.07 eV). These values are in good agreement with
the available measurements.

The  $\rm G_0W_0$ approximation evaluated using HSE orbitals 
 further  increases the band gap between the  occupied and empty states
resulting in significantly larger quasiparticle gaps than HSE (see Tab. \ref{data}). 
The update of the eigenvalues in G and, in particular, the inclusion of {\em e-h} 
(electron-hole) interactions ($\rm GW_0^{TCTC}$) leads to a slight
reduction of the quasiparticle band gaps ($\approx$ 4\%).
We note that the former approach ($\rm G_0W_0$)
is known to overestimate the band gaps for {\em s-p} bonded semiconductors and insulators using HSE
orbitals by typically
10 \%\cite{fuchs07,Paier08} whereas the latter approach ($\rm GW_0^{TCTC}$) yields very 
good values almost on par with fully self-consistent $\rm scQPGW^{TCTC}$ calculations\cite{Paier08,scgwvasp}.
Indeed, in this specific material, the $\rm scQPGW^{TCTC}$  band structure is very close to the 
$\rm GW_0^{TCTC}$ band structure, although the indirect band gap increases slightly from  1.15~eV to 1.28~eV. 
Unfortunately, the experimental uncertainty in $E_i$ does not permit to
ascertain which method yields the most accurate results, but  the GW values are generally
somewhat larger than the maximum values determined experimentally.
Since the direct gap $E_d$ has been measured using optical techniques, it is 
possibly impaired by excitonic effects (see below),
but overall it seems that GW does yield a slightly too large band gaps for
\bab.
However, compared to photoemission data\cite{peak}, the calculated
total density of states (shown on the right side of Fig.\ref{band}) show that 
$\rm GW_0^{TCTC}$ and $\rm scQPGW^{TCTC}$ yield a main valence band peak ($\approx$ -3.4 eV) in much better agreement with experiment (-3.2 eV)
than PBE (-1.9 eV) and HSE (-2.8 eV). Once more, we underline
the need of more detailed experimental data to 
unequivocally probe the predictive power of GW-type and HSE calculations. 
It should also be stressed that GW-like approaches, though employed for more than 20 years,
have been applied so far mainly to prototypical semiconductors and insulators (GaS, ZnO, $\rm Cu_2O$ etc...) but, 
to our knowledge, their performance on more complex systems such as \bab, has not yet been assessed in literature.
A comparison with accurate experimental measurements is therefore essential.

\begin{figure}
\includegraphics[clip,width=0.5\textwidth]{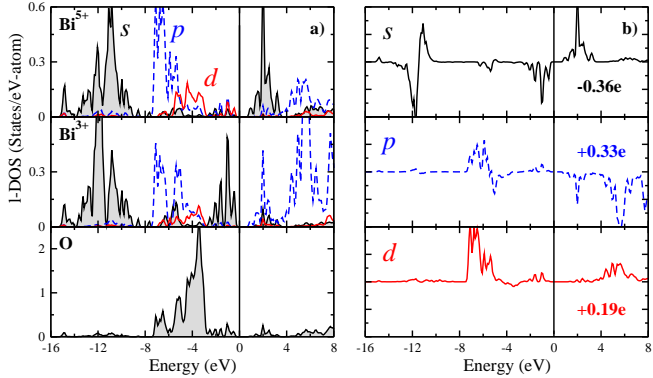}
\caption{(Color online) Calculated $\rm scQPGW^{TCTC}$ (a) partial DOS decomposed over Bi1, Bi2 and 
O (apical and planar Oxygen yield identical DOS) and (b) DOS difference between Bi1 and Bi2 
decomposed into $l$ quantum numbers.
}
\label{dos}
\end{figure}

\begin{table}[ht]
\caption{$\rm scQPGW^{TCTC}$, HSE and PBE  band-projected charge (in partial number of 
electrons $e^-$) on Bi1 and Bi2. Charges have been determined inside the PAW spheres.
}
\vspace{0.3cm}
\begin{ruledtabular}
\begin{tabular}{lccc}
             & $s$  & $p$ &  $d$  \\
             &      &     &       \\ 
             &\multicolumn{3}{c}{$\rm scQPGW^{TCTC}$}\\
  Bi1        &1.280 &1.199&10.254 \\
  Bi2        &1.642 &0.871&10.065 \\
  Bi1-Bi2    &-0.36 &+0.33&+0.19  \\
             &      &     &       \\ 
             &\multicolumn{3}{c}{HSE}\\
  Bi1        &1.318 &1.176&10.252 \\
  Bi2        &1.694 &0.830&10.063 \\
  Bi1-Bi2    &-0.38 &+0.35&+0.19  \\
             &      &     &       \\ 
             &\multicolumn{3}{c}{PBE}\\
  Bi1        &1.371 &1.085&10.181 \\
  Bi2        &1.623 &0.863&10.056 \\
  Bi1-Bi2    &-0.25 &+0.22&+0.13  \\
\end{tabular}
\end{ruledtabular}
\label{lcharge}
\end{table}

Let us proceed to shed light on the nature of the CDW gap $E_i$. It is accepted that
$E_i$ is formed by the splitting of the conduction Bi($s$)-O($p$) anti-bonding orbitals
into two subbands, but the electronic charge redistribution has not been discussed in details. 
Figure \ref{dos} (a) reports the $\rm scQPGW^{TCTC}$ {\em l}-projected density of states (DOS) 
on Bi and O sites which should be 
interpreted in conjunction with the band-projected charge (Tab. \ref{lcharge}). The formation of the CDW   
lowers the symmetry of the crystal thus inducing a global rearrangement of the charge around 
the two inequivalent Bi sites (see Fig. \ref{bt}).
We find that $E_i$ is opened between occupied  Bi2($s$) states (valence band maximum)
and empty Bi1($s$)-Bi2($p_xp_y$) orbitals (conduction band minimum).
This orbital-ordered insulating state is indeed more complex than the 
generally accepted picture invoking a simple Bi($s$)-O($p$) gap and arises from the elaborate
nature of the charge disproportionation, resulting in the electronic states shown in Figure \ref{dos} (b).

Previous calculations\cite{kunc,matth} argued that the static valence charge difference $\delta$
between Bi1 and Bi2 is not significant. 
The calculated $\delta$ is indeed rather small
($\rm \delta_{PBE} = 0.10e^-$, $\rm \delta_{HSE} = 0.16e^-$, $\rm \delta_{\rm scQPGW^{TCTC}} \approx 0.16e^-$) but it is the result of 
a substantial transfer involving orbitals with different angular moments. HSE and GW provide essentially the 
same picture, whereas PBE underestimates significantly the $l$-decomposed charge transfer by about 30\% (see Table \ref{lcharge}). 
In more detail, the flow of charge induced by the charge disproportionation between Bi1 and Bi2
involves a $s$-like charge transfer from Bi1 to Bi2 (0.36$e^-$) which is 
compensated by a backtransfer of {\em p} (0.33$e^-$) and {\em d} (0.16$e^-$) charge mostly 
localized in the energy range from -8 and -4 eV. This backtransfer can be  understood by realizing that
the O atoms are much closer to Bi1. As a result, the tails of the O($p$) orbitals overlap with the Bi1 spheres
and are picked up as states with higher angular momentum specifically $p$ and $d$ like character.

\subsection{Optical and dielectric properties}

\begin{figure}[hb]
\includegraphics[clip,width=0.45\textwidth]{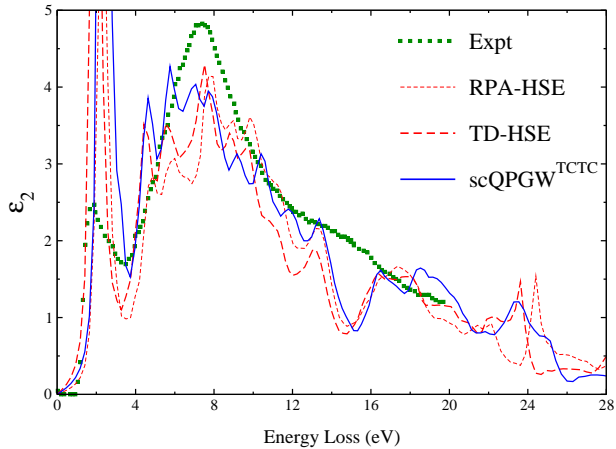}
\caption{(Color online) The imaginary parts of the dielectric function ($\epsilon_2$)
are shown for  RPA-HSE, TD-HSE  and $\rm scQPGW^{TCTC}$
along with  the experimental curve from Ref.\onlinecite{expdielectric}.
}
\label{eps}
\end{figure}

In addition to the formation of a band gap, the CDW also causes a 
strong optical resonance at $\approx$ 2 eV. In Fig.\ref{eps} we compare the experimental imaginary part 
of the dielectric function ${\epsilon_2}$\cite{expdielectric} with the theoretical spectrum 
computed within different approximations:
(i) the random phase approximation (RPA) using HSE orbitals, (ii) TD-HSE 
(as described in Ref. \onlinecite{Paier08}) and (iii) $\rm TD-scQPGW^{TCTC}$.
The first approach (i) neglects any electrostatic interactions between electrons and holes. The 
second approach (ii) is similar to TD Hartree-Fock but the electrostatic 
interaction between electrons and holes is only 1/4--- more precisely it is
described by the sr-exchange kernel in HSE ---whereas in
the final approach (iii) the interaction between electrons and holes is described
fully ab-initio using the W determined in the scQPGW calculations. For (i)
and (ii) HSE orbitals and one-electron energies are used, whereas (iii) is
based on the scQPGW one electron energies and orbitals, and the results of
(iii) are usually equivalent to Bethe-Salpeter calculations without the
Tamm-Dankof approximation.\cite{sottile,Paier08}

Considering the complexity of the system the agreement is remarkably good with regard to
the position and the width of the two dominant peaks 
namely the CDW peak at 2 eV and the excitation at 8.0 eV, which can be described as the transition from
the main peak of the valence band (-3.4 eV) to the broad group of states starting at about 4 eV
above $E_F$. The overestimation of the intensity of the CDW peak is most likely
related to a too coarse k-point sampling. Unfortunately, in this fairly complex system,
calculations with a denser k-point grid are very time consuming, and we
do not expect that they will change the results qualitatively. 
The first important observation is that there is a difference between the 
RPA-HSE and TD-HSE calculations. Excitonic effects (e-h interaction) shift
the CDW peak down by approximately 0.3-0.4 eV. The scQPGW calculations, which
also include excitonic effects, predict a blue shifted CDW excitation $\approx$ 0.2-0.4 eV
above the experimental peak. Unfortunately, this confirms that the 
scQPGW overestimates the direct band gap somewhat, as was already hinted at in
the section on the electronic properties.

Finally, we calculated the ion-clamped (high frequency) 
macroscopic dielectric constant ${\epsilon_\infty}$ (Tab.\ref{data}).
The HSE ${\epsilon_\infty}$ was computed adopting the perturbation expansion after discretization (PEAD) method\cite{pead}.
Within the PEAD method we applied a finite field in the supercell and extracted the static dielectric constant from
the resultant macroscopic field. This approach includes local field effects (electron-hole interactions). The so obtained macroscopic dielectric constant
(${\epsilon_\infty}^{HSE}$=6) is in good agreement with the experimental estimations (5 and 7, see Tab.\ref{data})
extracted through a fit of the reflectivity spectra.
The $\rm GW_0$ and scQPGW values, 7.0 and 7.2 respectively, though still in good agreement with 
experiment are about 15\% larger than the HSE values.
As expected from the serious underestimation of the gap, TD-PBE fails 
dramatically in predicting ${\epsilon_\infty}$ (${\epsilon_\infty}^{\rm PBE}$=12.2).

\subsection {Phonons and infrared reflectivity} 

\begin{figure}
\includegraphics[clip,width=0.48\textwidth]{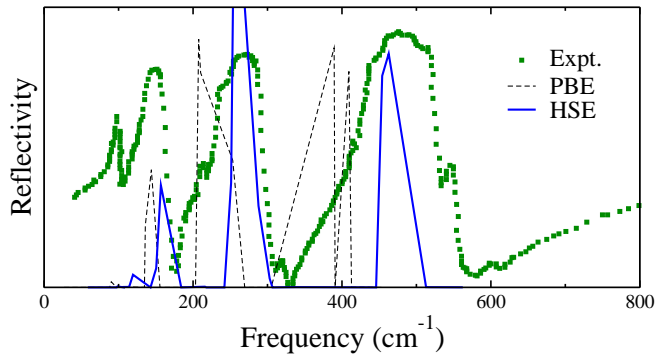}
\caption{(Color online) Comparison between calculated and measured\cite{reflectivity} infrared reflectivity.}
\label{ph}
\end{figure}

We conclude our analysis by exploring the effect of non-local exchange on the zone-centered phonon
frequencies and on the oscillator strengths of the infrared reflectivity (IR) active modes. 
Since within the GW method, the calculation of phonon frequencies is presently out of reach, we
focus on PBE and HSE only.
The evaluation of the Born effective charge (BEC) $Z^{*}_{i,\alpha{\beta}}$, required for 
the calculation of the IR spectra, were performed adopting the PEAD (HSE) and linear response (PBE) methods.
Excluding the three acoustic modes (which are zero at $\Gamma$), 
the monoclinic phase (containing 10 atoms per unit cell) possesses 27 optical modes with the symmetry
\begin{displaymath} 7A_g + 5B_g + 6A_u + 9B_u \end{displaymath}. The {\em gerade} modes   
are Raman active and the {\em ungerade} modes are infrared active. In Fig. \ref{ph} we compare our
calculated HSE IR spectra with the experimental reflectivity data\cite{reflectivity}. 
The results are remarkable: the inclusion of non-local exchange is essential to account for
a correct prediction of the phonon modes with large oscillator strength.
The three highest modes at 157, 256 and 463 $\rm cm^{-1}$ reproduce very well the corresponding experimental values, 
whereas the lowest mode at 119 $\rm cm^{-1}$ is shifted upward with respect to the 
measured one by $\approx$ 20 $\rm cm^{-1}$.
In addition, we find that the highest frequency phonon (Raman active)
corresponding to the $\rm BiO_6$ breathing mode responsible for the CDW instability, 
experiences a huge shift from PBE (453 $\rm cm^{-1}$) to HSE (562 $\rm cm^{-1}$). The HSE value coincides
with the value provided by Raman scattering analysis\cite{raman}, 570 $\rm cm^{-1}$. 
The problems  of PBE can be attributed primarily to volume effects: Due to the large PBE volume 
the PBE phonon vibrations are located at lower frequencies. In addition to this, we note that
the PBE inadequate description of the oxygen breathing distortions is also reflected in the phonon spectra: 
The frequency downshift is larger (45-60 $\rm cm^{-1}$) for the oxygen-based modes
lying at higher frequency than for the lower modes (20-30 $\rm cm^{-1}$) emerging from the softer Bi and Ba 
vibrations which are actually very well describe at the PBE level.
A similar behaviour has been observed in ZnO\cite{hummer}

\begin{table}[h]
\caption{Calculated HSE and PBE Born effective charges $Z^{*}_{i,\alpha{\beta}}$ (with $i$
atom index) in unit of $\lvert{e^-}\rvert$. $Z^{*}_{\parallel}$ and $Z^{*}_{\perp}$ refers to the components
parallel and perpendicular to the Bi-O bond. The LDA values are taken from Ref. \onlinecite{rabe}.
}
\vspace{0.3cm}
\begin{ruledtabular}
\begin{tabular}{lcccc}
                           & Bi1  & Bi2  &  Ba   & O \\
                           &      &      &       &   \\
                           &\multicolumn{4}{c}{HSE} \\
$Z^{*}_{i,\alpha{\alpha}}$ & 5.01 & 5.98 &  2.71 &   \\
$Z^{*}_{\parallel}$        &      &      &       & -4.39  \\
$Z^{*}_{\perp}$            &      &      &       & -1.88  \\
                           &      &      &       &   \\
                           &\multicolumn{4}{c}{PBE} \\
$Z^{*}_{i,\alpha{\alpha}}$ & 4.65 & 6.31 &  2.71 &   \\
$Z^{*}_{\parallel}$        &      &      &       & -4.36  \\
$Z^{*}_{\perp}$            &      &      &       & -1.89  \\
                           &      &      &       &   \\
                           &\multicolumn{4}{c}{LDA\cite{rabe}} \\
$Z^{*}_{i,\alpha{\alpha}}$ & 4.78 & 6.22 &  2.75 &   \\
$Z^{*}_{\parallel}$        &      &      &       & -4.55  \\
$Z^{*}_{\perp}$            &      &      &       & -1.85  \\

\end{tabular}
\end{ruledtabular}
\label{z}
\end{table}

Finally, we list in Tab. \ref{z} the computed BEC $Z^{*}_{i,\alpha{\beta}}$. 
We find that the BEC tensor is diagonal and reduces to a scalar matrix for Ba and Bi
(i.e. $Z^{*}_{Ba/Bi,11} = Z^{*}_{Ba/Bi,22} = Z^{*}_{Ba/Bi,33}$) and to two
components $Z^{*}_{\parallel}$ and $Z^{*}_{\perp}$ for O with respect to the Bi-O bond.
In agreement with precedent estimations of Thonhauser and Rabe\cite{rabe} we observe a
significant BEC disproportionation between Bi1 and B2 ($Z^{*}_{Bi2}-Z^{*}_{Bi1}$),
substantially larger than the corresponding static charge transfer $\delta$ discussed 
above (see Tab. \ref{lcharge}). The incorporation of non-local exchange effects through HSE
reduces the dynamical charge disproportionation from 1.66 $\lvert{e^-}\rvert$ (PBE) to
0.97 $\lvert{e^-}\rvert$ (HSE) and leaves the Ba and O BEC's unchanged. 
The average BEC of oxygen -3.13 $\lvert{e^-}\rvert$ equals the experimentally 
derived value of -3.2 $\lvert{e^-}\rvert$\cite{expeps2}.

%================================
\section{Summary}
\label{sec:4}
%===============================

In summary, by means of beyond-DFT methods we have computed ground and excited state properties of \bab.
On the basis of hybrid functionals and fully {\em ab initio} GW methods, we have shown that the inclusion of
non-local exchange effects is essential to reproduce and understand the complex physical properties of \bab.                            
Our calculated structural distortions (oxygen instabilities and cubic-to-monoclinic transition), 
optical excitations (direct/indirect band gap and optical spectrum), 
dielectric (dielectric constant) and 
vibrational properties (phonon frequencies, infrared reflectivity and Born effective charges) 
reproduce well the available experimental findings. 
Overall, our study suggests that the GW approaches tend to overestimate
the band gap, which seems to be better predicted within the screened hybrid HSE functional.
Although the inclusion of an effective non-local exchange-correlation kernel $f_{xc}$
at TD-HSE ($\rm GW_0^{TCTC}$) and BSE-GW ($\rm scQPGW^{TCTC}$) level improves the single-shot $G_0W_0$ 
description, the resulting band gap is
still $\approx$ 15\% (E$_d$) and 40\% (E$_i$) larger than the corresponding {\em bare} HSE value.
The source of this error as far as it can not be attributed to uncertainties in the experimental results
likely resides in the neglect of many-body vertex corrections in the self-energy $\Sigma$, which are entirely
neglected in this work, or any other GW implementation.
Though $\rm scQPGW^{TCTC}$ has proven to be excellent for
the prediction of the band gaps of typical semiconductor, insulator and noble gas solids it was
never applied to more complex oxides such as $\rm BaBiO_3$. 
Given the lack of accurate experimental information on the electronic properties
of $\rm BaBiO_3$ (especially for the indirect gap E$_i$) the assessment of these GW techniques to 
less critical oxides for which well established experimental data exists is highly required.  

Concluding, our research on $\rm BaBiO_3$ clarifies the serious drawbacks of
standard DFT and provide a key to understand and design similar charge-ordered materials.
We hope that our study can boost for new experiments as well as for thorough theoretical 
and computational efforts aiming to straighten the issues still under debate.
Furthermore, we have shown that the hybrid  functional HSE yields a description which is 
at least on par with sophisticated man-body techniques.

\end{document}